\documentclass[11pt, a4paper]{article}
\usepackage{jheppub}
\usepackage{amssymb,amsfonts,amsmath,subfigure}
\usepackage{graphicx,epsfig,pdfpages}
\newcommand{\bea}{\begin{eqnarray}}
\newcommand{\eea}{\end{eqnarray}}
\def\gsim{\mathrel{
   \rlap{\raise 0.511ex \hbox{$>$}}{\lower 0.511ex \hbox{$\sim$}}}}
\def\lsim{\mathrel{
   \rlap{\raise 0.511ex \hbox{$<$}}{\lower 0.511ex \hbox{$\sim$}}}}

\title{7 keV Sterile neutrino dark matter in \boldmath{$U(1)_{R^-}$}lepton 
number model}

\author[1]{Sabyasachi Chakraborty {\note{Corresponding author}},}
\author{Dilip Kumar Ghosh}
\author{and Sourov Roy}

\affiliation{Department of Theoretical Physics, Indian Association for the 
Cultivation of Science, 2A $\&$ 2B Raja S.C.Mullick Road, Jadavpur, 
Kolkata 700 032, INDIA.}

\emailAdd{tpsc3@iacs.res.in}
\emailAdd{tpdkg@iacs.res.in}
\emailAdd{tpsr@iacs.res.in}

\abstract{
We study the phenomenology of a keV sterile neutrino in a supersymmetric model 
with $U(1)_{R^-}$lepton number in the light of a very recent observation of an 
X-ray line signal at around 3.5 keV, detected in the X-ray spectra of Andromeda 
galaxy and various galaxy clusters including the Perseus galaxy cluster. This 
model not only provides a small tree level mass to one of the active neutrinos 
but also renders a suitable warm dark matter candidate in the form of a sterile 
neutrino with negligible active-sterile mixing. Light neutrino masses and mixing 
can be explained once one-loop radiative corrections are taken into account. 
The scalar sector of this model can accommodate a Higgs boson with a mass of 
$\sim$ 125 GeV. In this model gravitino is the lightest supersymmetric particle 
(LSP) and we also study the cosmological implications of this light gravitino 
with mass $\sim \mathcal O$(GeV).}

\keywords{Supersymmetry Phenomenology}
\arxivnumber{1405.6967}

\begin{document}
\maketitle
\flushbottom
\section{Introduction}
We are living in an era enriched with many experimental breakthroughs and 
results especially in the area of astro-particle physics and cosmology. 
The most recent one is the identification of a weak line at $E\sim 3.5 
~\rm{keV}$ in the X-ray spectra of the Andromeda galaxy and many other galaxy 
clusters including the Perseus galaxy cluster, observed by XMM-Newton X-ray 
Space observatory\cite{Bulbul, Boyarsky}. The observed flux and the best fit 
energy peak are at
\bea
\Phi_{\gamma}&=&4\pm 0.8\times 10^{-6} 
~\rm{photons~cm^{-2} sec^{-1}},\nonumber \\
E_{\gamma}&=&3.57\pm 0.02 ~\rm{keV}.
\eea
Since atomic transitions in thermal plasma cannot account for this energy, 
therefore the concept of a dark matter, providing the possible explanation 
regarding the appearance of this photon line becomes extremely important. 
This result can be explained by a sterile neutrino \cite{Ishida, Abazajian, 
Modak, Cline, Barry, Robinson}, axion or axion like warm dark matter 
\cite{Higaki,Jaeckel,Lee,Conlon}, axino \cite{Kong,Choi,Liew}, excited dark
matter \cite{Okada,Okada-1}, gravitino \cite{Bomark,Demidov}
and keV scale LSP \cite{Kolda} as decaying dark matter. Other interesting
scenarios with an annihilating scalar dark matter \cite{Dudas}, decaying
Majoron \cite{Sinha} and a keV scale dark gaugino \cite{Kang} have also 
been considered in this context. In this work we consider sterile neutrino in a 
$U(1)_{R^-}$lepton number model, which could provide a possible explanation for 
the emergence of the photon line. The observed flux and the peak of the 
energy readily translates to an active-sterile mixing in the range $2.2 
\times 10^{-11}<\sin^2 2\theta_{14} < 2 \times 10^{-10}$ and the mass of the 
sterile neutrino dark matter $M^R_N = 7.06 \pm 0.05$ keV \cite{Boyarsky}.

On the other hand, in high energy collider frontier two CERN based 
experiments ATLAS and CMS have confirmed the existence of a neutral 
elementary scalar boson of nature, with mass around 125 GeV 
\cite{Aad,Chatrchyan}. Nevertheless, more analysis is required to confirm it as 
the Standard Model (SM) Higgs boson. In order to explain the mass of this 
scalar boson in a natural way, to address the question of nonzero neutrino mass 
and mixing and to provide a candidate for dark matter, many beyond standard 
model (BSM) theories have been pursued for quite some time and supersymmetry 
remains one of the most celebrated ones as of now. However, supersymmetric 
particle searches by ATLAS and CMS experiments for pp collision at center of 
mass energy 7 and 8 TeV, have observed no significant excess 
\cite{Chatrchyan-1, Aad-1} over the standard model background. This has 
put very stringent lower limits on the superpartner masses. 

In the light of this present situation, $U(1)_{R^-}$symmetric models with Dirac 
gauginos are well motivated because they can relax the strong bounds on the 
superpartner masses, explain the 125 GeV Higgs boson mass, provide non-zero 
neutrino mass at the tree as well as at the one loop level and can also 
accommodate a suitable dark matter candidate. Various aspects of different 
R-symmetric models have been studied and can be found in the literature 
\cite{Fayet,Hall-1,Hall-2, Nelson,Fox-1,Chacko,Choi-S,Kribs,Benakli,Kumar,Fox,
Benakli-2,Benakli-1,Davies,Davies-1,Gregoire,Greg,Bertuzzo,Goodsell,Riva,Kumar-1,Claudia,
Chakraborty,Dudas-1,Beauchesne,Benakli:2014cia}. In this work, we study a 
particular $U(1)_{R^-}$symmetric model where we have identified the R-charges 
with lepton numbers in such a way that the lepton numbers of the standard model 
fermions correspond to the negative of their R-charges \cite{Kumar-1,Claudia}. 
The role of the down-type Higgs is played by the sneutrino since its vacuum 
expectation value (vev) is not constrained by the Majorana mass of the neutrino. 
The minimal extension of this model by adding a single right handed neutrino 
superfield also gives rise to very interesting phenomenological consequences 
\cite{Chakraborty}. It generates a tree level Dirac mass for one of the 
neutrinos in the R-symmetry preserving scenario. If R-symmetry is broken because 
of the presence of a non zero gravitino mass, then for small neutrino Yukawa 
coupling, $f\sim\mathcal O(10^{-4})$, the extended neutralino-neutrino mass 
matrix provides a sterile neutrino state accompanied by an active neutrino 
state. Here we identify the sterile neutrino as the warm dark matter in our 
model.

The presence of R-symmetry inhibits gauginos to acquire a Majorana mass. 
However, gauginos can have Dirac masses and to introduce the Dirac gaugino 
mass, one must consider a singlet chiral superfield $\hat S$, a triplet 
$\hat T$ and an octet $\hat O$ living in the adjoint representation of 
$U(1)_Y$, $SU(2)_L$ and $SU(3)_C$ respectively. The Dirac gaugino masses 
are also coined as `supersoft' mass terms since they do not contribute 
to any logarithmic corrections to the scalar masses. The presence of Dirac 
gluino also helps to relax the bound on squark masses compared to MSSM and 
in addition flavor and CP violating constraints are suppressed in this class 
of models \cite{Kribs}.

The plan of the paper is as follows. At first we describe the model in section 
II, with appropriate R-charge assignments. In section III we discuss very 
briefly, the scalar sector of the model and point out the extra contributions 
to the Higgs boson mass, which can arise both at the tree level as well as 
at the one loop level. Section IV addresses the issue of R-symmetry breaking 
and tree level Majorana masses of the sterile and one of the active neutrinos.
Next in section V the essential features of the sterile neutrino as a keV 
warm dark matter candidate are discussed and its production mechanism and the 
dominant decay modes relevant to our model are highlighted. In section VI we 
briefly present a discussion related to the cosmology of the gravitino in 
this model with a few GeV mass and finally, in section VII, we summarise 
our results.
  
\section{$U(1)_{R^-}$lepton number model with a right handed neutrino superfield}

We study a $U(1)_{R^-}$lepton number model, where in addition to the standard 
superfields of the MSSM - $\hat H_u$, $\hat H_d$, $\hat Q_i$, $\hat {U_i}^c$, 
$\hat {D_i}^c$, $\hat L_i$, $\hat {E_i}^c$, this model includes a right handed 
neutrino superfield and a pair of vector-like $SU(2)_L$ doublet superfields 
$\hat R_u$ and $\hat R_d$, with opposite hypercharge \cite{Chakraborty}. These two 
doublets carry non zero R-charges (The R-charge assignments are given in Table I) 
and therefore, to avoid spontaneous R-breaking and the emergence of R-axions, they
do not acquire any non-zero vev and would remain inert. R symmetry 
prohibits soft supersymmetry breaking terms like Majorana gaugino masses and 
trilinear scalar couplings. However, gauginos can acquire Dirac masses as 
mentioned in the introduction. The implications of adding a right-handed 
neutrino superfield $\hat N^c$ is discussed later in detail. We would like
to reiterate that the R-charge assignments are such that the lepton number of 
the SM fermions are negative of their corresponding R-charges. 

\begin{table}[h!]
\begin{center}
\begin{tabular}{|c|ccccccccccccc|}
\hline
\rule{0mm}{5mm}
& $\hat Q_{i}$ & $\hat U_{i}^{c}$ & $\hat D_{i}^{c}$ & $\hat L_{i}$
& $\hat E_{i}^{c}$ & $\hat H_{u}$ & $\hat H_{d}$ & $\hat R_{u}$
& $\hat R_{d}$ & $\hat S$ & $\hat T$ & $\hat O$ & $\hat N^{c}$
\\[0.3em]
\hline
\rule{0mm}{5mm}
$U(1)_{R}$ & 1 & 1 & 1 & 0 & 2 & 0 & 0 & 2 & 2 & 0 & 0 & 0 & 2 \\
[0.3em]
\hline
\end{tabular}
\end{center}
\vspace{-10pt}
\caption{$U(1)_{R}$ charge assignments of the chiral superfields.}
\label{R-charges}
\vspace{-15pt}
\end{table}
The generic superpotential, carrying R-charge of 2 units is
\bea
W&=&y^{u}_{ij}\hat H_{u}\hat Q_{i}\hat U^{c}_{j}
+\mu_{u}\hat H_{u}\hat R_{d}+f_{i}\hat L_{i}\hat H_{u}\hat N^{c}
+\lambda_{S}\hat S\hat H_{u}\hat R_{d} 
+2\lambda_{T}
\hat H_{u}\hat T\hat R_{d}-M_{R}\hat N^{c}\hat S+
\mu_{d}\hat R_{u}\hat H_{d}\nonumber \\
&+&\lambda^\prime_{S}\hat S\hat R_{u}\hat H_{d} 
+\frac{1}{2}\lambda_{ijk}
\hat L_{i}\hat L_{j}\hat E^{c}_{k}
+\lambda^{\prime}_{ijk}\hat L_{i}\hat Q_{j}\hat D^{c}_{k}
+2\lambda^\prime_{T}\hat R_{u}\hat T\hat H_{d} 
+y^{d}_{ij}\hat H_{d}
\hat Q_{i}\hat D^{c}_{j}+y^{l}_{ij}\hat H_{d}
\hat L_{i}\hat E^{c}_{j} \nonumber \\
&+& \lambda_N {\hat N}^c {\hat H}_u {\hat H}_d.
\label{superpotential}
\eea
Note that a subset ($\lambda$, $\lambda^{\prime}$) of standard R-parity violating 
operators are present in the superpotential although the model is $U(1)_R$ 
conserving (i.e. lepton number conserving). In a somewhat simplistic approach 
we have omitted the terms $\hat N^c \hat S\hat S$ and $\hat N^c$ from the 
superpotential. 

In a realistic model one should also include supersymmetry breaking terms, 
such as the gaugino and scalar mass terms. The Dirac gaugino `supersoft' mass 
terms are constructed from a spurion superfield $W^{\prime}_{\alpha} = 
\lambda_{\alpha}+\theta_{\alpha}D^{\prime}$, if supersymmetry breaking is of 
the D-type. The Lagrangian containing the Dirac gaugino masses are 
\cite{Benakli-2,Benakli-1} 
\bea
{\cal L}^{\rm Dirac}_{\rm gaugino} &=& \int d^2 \theta 
\dfrac{W^\prime_\alpha}{\Lambda}[\sqrt{2} \kappa_1 ~W_{1 \alpha} {\hat S} 
+ 2\sqrt{2} \kappa_2 ~{\rm tr}(W_{2\alpha} {\hat T}) 
+ 2\sqrt{2} \kappa_3 ~{\rm tr}(W_{3\alpha} {\hat O})] + h.c.
\label{dirac-gaugino}
\eea
\vspace {2mm}
This D-term breaking generates Dirac mass for the gauginos, proportional 
to $k_i\frac{<D^{\prime}>}{\Lambda}$, where $\Lambda$ denotes the scale of 
SUSY breaking mediation. In a similar manner the $U(1)_R$ conserving soft 
supersymmetry breaking terms in the scalar sector are generated by the spurion 
superfield $\hat X$, defined as $\hat X=x+\theta^2 F_X$. The non-zero 
vev of the F-term generates the scalar soft terms as
\bea
V_{soft}&=& m^{2}_{H_{u}} H_{u}^{\dagger}H_{u}+m^{2}_{R_{u}}
R_{u}^{\dagger}R_{u}+ m^{2}_{H_{d}}H_{d}^{\dagger}H_{d}
+m^{2}_{R_{d}}R_{d}^{\dagger}R_{d}
+m^{2}_{\tilde L_{i}}\tilde L_{i}^{\dagger}\tilde L_{i}\nonumber \\
&+&m^2_{{\tilde R}_i}{{\tilde l}^\dagger_{Ri} {\tilde l}_{Ri}} +
M_{N}^{2}\tilde N^{c\dagger}\tilde N^{c}
+m_{S}^{2} S^{\dagger}S+2m_{T}^{2} {\rm tr}(T^{\dagger}T)
+2 m_O^2 {\rm tr}(O^\dagger O) \nonumber \\
&+& (B\mu H_u H_d + {\rm h.c.})- (b\mu_L^i H_u {\tilde L}_i + {\rm h.c.}) 
+(t_{S}S+{\rm h.c.})\nonumber \\
&+&\frac{1}{2} b_{S}(S^{2}+{\rm h.c.})
+b_{T} ({\rm tr}(TT) + {\rm h.c.})+B_O({\rm tr}(OO) + {\rm h.c.}).
\label{soft-scalar-terms}
\eea
The presence of the bilinear term $b\mu_L^i H_u\tilde L_i$ in the soft 
supersymmetry breaking potential implies all the three left handed sneutrinos 
can acquire a non zero vevs ($v_i$). To simplify, we perform a basis rotation as 
$\hat L_i=\frac{v_i}{v_a}\hat L_a+e_{ib}\hat L_b$ by which only one of the 
sneutrinos acquire a non zero vev ($v_a$) and we choose it to be the electron 
sneutrino ($a=1(e)$). We also choose the neutrino Yukawa coupling ($f$) in such a 
manner that only $\hat L_a$ couples with $\hat N^c$, the right-handed neutrino 
superfield \cite{Chakraborty}. Finally, we choose a very large $\mu_d$ such that 
the superfields $\hat H_d$ and $\hat R_u$ gets decoupled, which also implies 
that the left handed electron type sneutrino now plays the role of a down type 
Higgs field. We would like to emphasise that the model is lepton number 
conserving and therefore, the sneutrino vev is not constrained from the Majorana 
mass of the neutrinos. This is clearly different from the standard R-parity 
violating scenario.

In the mass eigenstate basis (primed superfields) of the down-type quarks and the 
charged leptons\footnote{Note that the mass of the charged lepton of flavor $a$
can come from R-symmetry preserving supersymmetry breaking operators \cite{Kumar-1}.}
the superpotential takes the following form \cite{Chakraborty}

\bea
W&=&y_{ij}^{u}\hat H_{u}\hat Q_{i}\hat U_{j}^{c}+\mu_{u}\hat 
H_{u}\hat R_{d}+f\hat L_{a}\hat H_{u}\hat N^{c}+
\lambda_{S}\hat S\hat H_{u}\hat R_{d} 
+2\lambda_{T}\hat H_{u}
\hat T\hat R_{d}-M_{R}\hat N^{c}\hat S + W^{\prime},\nonumber \\
\label{final-superpotential}
\eea
and
\bea
W^{\prime}&=&\sum_{b=2,3} f^l_b {\hat L^{\prime}}_a {\hat L^\prime}_b 
{\hat E^{\prime c}}_b + \sum_{k=1,2,3} f^d_k {\hat L^\prime}_a
{\hat Q^\prime}_k {\hat D^{\prime c}}_k 
+ \sum_{k=1,2,3} \dfrac{1}{2} 
{\tilde \lambda}_{23k}{\hat L^\prime}_2 {\hat L^\prime}_3 
{\hat E^{\prime c}}_k \nonumber \\
&+& \sum_{j,k=1,2,3;b=2,3}{\tilde \lambda}^\prime_{bjk}
{\hat L^\prime}_b {\hat Q^\prime}_j {\hat D^{\prime c}}_k. 
\label{W-diag}
\eea
In our subsequent analysis we stay in this mass basis
but remove the prime from the fields and make the replacement $\tilde\lambda$, 
$\tilde\lambda^{\prime}\rightarrow \lambda$, $\lambda^{\prime}$.
The soft supersymmetry breaking but $U(1)_R$ preserving terms in 
the rotated basis are
\bea
V_{soft}&=& m^{2}_{H_{u}}H_{u}^{\dagger}H_{u}+m^{2}_{R_{d}}
R_{d}^{\dagger}R_{d}+m^{2}_{\tilde L_{a}} \tilde L_{a}^{\dagger}
\tilde L_{a} 
+\sum_{b=2,3} m^{2}_{\tilde L_{b}} \tilde L_{b}^{\dagger}
{\tilde L_{b}}+M_{N}^{2}{\tilde N}^{c\dagger} {\tilde N}^{c}
+m^2_{{\tilde R}_i}{{\tilde l}^\dagger_{Ri} {\tilde l}_{Ri}} \nonumber \\
&+&m_{S}^{2} S^{\dagger}S+2m_{T}^{2} {\rm tr}(T^{\dagger}T) 
+2m_O^2 {\rm tr}(O^\dagger O) 
- (b\mu_L H_u {\tilde L}_a + {\rm h.c.}) 
+(t_{S}S+{\rm h.c.}) \nonumber \\
&+&\frac{1}{2} b_{S}(S^{2}+{\rm h.c.})
+b_{T} ({\rm tr}(TT) + {\rm h.c.}) +B_O({\rm tr}(OO) + {\rm h.c.}).
\label{final-softsusy-terms} 
\eea

In the R-symmetric case, the lightest eigenvalue of the neutralino mass matrix, 
written in the basis ($\tilde b^0$, $\tilde w^0$, $\tilde R_d^0$, $N^c$) and 
($\tilde S$, $\tilde T^0$, $\tilde H_u^0$, $\nu_e$), provides a tree level 
Dirac neutrino mass, which can be written as \cite{Chakraborty} 
\bea
m_{\nu_e}^D&=&\frac{v^3 \sin\beta f g \lambda_T}{\sqrt 2\gamma 
M_1^D M_2^D}(M_2^D-M_1^D),
\label{Dirac-neutrino} 
\eea
where $M_1^D$, $M_2^D$ stands for Dirac bino and wino masses respectively, 
$\gamma = \mu_u + \lambda_S v_S + \lambda_T v_T$, $g$ is the $SU(2)$ gauge 
coupling, $\tan\beta = \frac{v_u}{v_a}$, $v\equiv \sqrt{v_u^2+v_a^2} = 
\frac{\sqrt 2 M_W}{g}$. 
To obtain this particular form in eq.~(\ref{Dirac-neutrino}) we have assumed 
certain relations involving the parameters and they are
\bea
\lambda_T &=& \tan\theta_W\lambda_S, \nonumber \\
M_R &=& \frac{\sqrt 2 f M_1^D \tan\beta}{g\tan\theta_W}.
\label{MR-lambda} 
\eea
Therefore, with appropriate choice of parameters one can easily obtain a small 
tree level Dirac neutrino mass $\sim 0.1$ eV. 
\section{Scalar sector}
In this section we shall mention very briefly about the scalar sector of this 
particular model. For a detailed discussion we refer the reader to 
\cite{Chakraborty}. The lightest CP even scalar mass matrix, in the basis of 
($H_u$, $\tilde \nu$, $S$, $T$), provide the CP even Higgs boson. It is 
remarkable that the neutrino Yukawa coupling $f$ renders a tree level correction 
to the lightest Higgs boson mass, which we calculate as
\bea
M_h^2 \leq M_z^2 \cos^2 2\beta + f^2 v^2 \sin^2 2\beta.
\eea
For $f\sim\mathcal O(1)$ and for small $\tan\beta$, the tree level
\footnote{In this paper we shall not explore such a possibility and
concentrate on the region of parameter space where $f\sim \mathcal O(10^{-4})$,
which produces a keV sterile neutrino state.}
Higgs boson mass can satisfy the present observed value, close to $125$ GeV
\cite{Chakraborty}. It is also pertinent to mention that the singlet and the 
triplet fields provide very important loop corrections to the Higgs boson mass. 
These contributions can be sizable if the singlet and the triplet couplings
$\lambda_S$ and $\lambda_T$ are large. The dominant radiative corrections to 
the quartic potential can be written as \cite{Belanger}, $\frac{1}{2}\delta
\lambda_u(|H_u|^2)^2$, $\frac{1}{2}\delta\lambda_\nu(|\tilde\nu_a|^2)^2$ and 
$\frac{1}{2}\delta\lambda_3|H_u^0|^2|\tilde\nu_a|^2$, where
\bea
\delta\lambda_{u}&=& \frac{3 y_{t}^{4}}{16\pi^{2}} 
\ln \left(\frac{m_{\tilde t_{1}}m_{\tilde t_{2}}}{m_{t}^2}\right)
+\frac{5\lambda_{T}^{4}}{16\pi^{2}}\ln\left(\frac{m_{T}^{2}}
{v^{2}}\right) 
+\frac{\lambda_{S}^{4}}{16\pi^{2}}\ln\left(\frac{m_{S}^2}{v^{2}}\right)
-\frac{1}{16\pi^{2}}\frac{\lambda_{S}^{2}\lambda_{T}^{2}}
{m_{T}^{2}-m_{S}^{2}}\nonumber \\
&&\Big(m_{T}^{2}\Big\{\ln\left(\frac{m_{T}^{2}}{v^{2}}\right)-1\Big\}
-m_{S}^{2}\Big\{\ln \left(\frac{m_{S}^{2}}{v^{2}}\right)-1\Big\}\Big),\nonumber \\
\eea
\bea
\delta\lambda_{\nu}&=& \frac{3 y_{b}^{4}}{16\pi^{2}} 
\ln \left(\frac{m_{\tilde b_{1}}m_{\tilde b_{2}}}{m_{b}^2}\right)
+\frac{5\lambda_{T}^{4}}{16\pi^{2}}\ln\left(\frac{m_{T}^{2}}
{v^{2}}\right)
+\frac{\lambda_{S}^{4}}{16\pi^{2}}
\ln\left(\frac{m_{S}^2}{v^{2}}\right)
- \frac{1}{16\pi^{2}}\frac{\lambda_{S}^{2}\lambda_{T}^{2}}
{m_{T}^{2}-m_{S}^{2}}\nonumber \\
&&\Big(m_{T}^{2}\Big\{\ln\left(\frac{m_{T}^{2}}{v^{2}}\right)-1\Big\}
-m_{S}^{2}\Big\{\ln \left(\frac{m_{S}^{2}}{v^{2}}\right)-1\Big\}\Big), \nonumber \\
\eea
and finally,
\bea
\delta\lambda_{3}&=& \frac{5 \lambda_{T}^{4}}{32\pi^{2}} 
\ln (\frac{m_{T}^{2}}{v^{2}})
+\frac{1}{32\pi^{2}}\lambda_{S}^{4}
\ln\left(\frac{m_{S}^{2}}{v^{2}}\right)
+\frac{1}{32\pi^{2}}\frac{\lambda_{S}^{2}\lambda_{T}^{2}}{m_{T}^{2}
-m_{S}^{2}}\nonumber \\
&&\Big(m_{T}^{2}\Big\{\ln \left(\frac{m_{T}^{2}}{v^{2}}\right)-1\Big\}
-m_{S}^{2}\Big\{\ln\left(\frac{m_{S}^{2}}{v^{2}}\right)-1\Big\}\Big).\nonumber \\
\eea

Therefore, for large $\lambda_S$, $\lambda_T\sim\mathcal O(1)$, a
125 GeV Higgs boson mass can easily be accommodated in this model even in
the presence of a light stop mass and negligible left-right mixing.
\section{R-symmetry breaking}
Until now we have constrained ourselves in the R-symmetry preserving
scenario. Although the R-symmetric case in this regard is interesting and 
should be explored in much more detail but in our work we pursue the path, 
where R-symmetry is broken. Recent cosmological observations point towards 
a vanishingly small vacuum energy or cosmological constant associated with 
our universe. Spontaneously broken supergravity theory in a hidden sector 
requires a non zero value of the superpotential in vacuum in order to have 
this small vacuum energy. As the superpotential carries R-charge of two units 
($R[W]=2$), therefore R-symmetry is broken when the superpotential acquires 
a non zero vev $\langle W \rangle$. Furthermore, a non zero gravitino 
mass also requires a non zero $\langle W \rangle$, thereby one can consider 
the gravitino mass as the order parameter of R-symmetry breaking.

The breaking of R-symmetry has to be communicated to the visible sector and 
in this context we confine ourselves to the case of anomaly mediation, 
which plays the role of the messenger of R-symmetry breaking \cite{Kumar-1, 
Chakraborty}. Such a scenario generates very small ($\sim$ a few MeV) Majorana 
gaugino masses and trilinear scalar couplings, $M_i\sim \frac{g_i^2}
{16\pi^2}m_{3/2}$ and $A_{u/d}=\frac{\hat\beta_{h_{u/d}} v_{u/d}}{16\pi^2 
m_{u/d}} m_{3/2}$ \cite{Guidice,Ghosh}, as long as the gravitino mass is in 
the range of a few GeV.

In the R-breaking case, the neutralino mass matrix written in the 
basis \\
$(\tilde b^{0}, \tilde S, \tilde w^{0}, \tilde T, \tilde R_{d}^{0}, 
\tilde H_{u}^{0}, N^{c}, \nu_{e})$, is given by
\bea
M_{\chi}^{M}=\left(
\begin{array}{cccccccc}
M_{1} & M_{1}^{D} & 0 & 0 & 0 & \frac{g^{\prime}v_{u}}{\sqrt 2} & 0 
& -\frac{g^{\prime}v_{a}}{\sqrt 2}\\
M_{1}^{D} & 0 & 0 & 0 & \lambda_{S}v_{u} & 0 & M_{R} & 0\\
0 & 0 & M_{2} & M_{2}^{D} & 0 & -\frac{g v_{u}}{\sqrt 2} & 0 & 
\frac{g v_{a}}{\sqrt 2}\\
0 & 0 & M_{2}^{D} & 0 & \lambda_{T}v_{u} & 0 & 0 & 0 \\
0 & \lambda_{S}v_{u} & 0 & \lambda_{T}v_{u} & 0 & \mu_{u}+
\lambda_{S}v_{S}+\lambda_{T}v_{T} & 0 & 0\\
\frac{g^{\prime}v_{u}}{\sqrt 2} & 0 & -\frac{gv_{u}}{\sqrt 2}& 0 &
\mu_{u}+\lambda_{S}v_{S}+\lambda_{T}v_{T} & 0 & -fv_{a} & 0\\
0 & M_{R} & 0 & 0 & 0 & -fv_{a} & 0 & -fv_{u} \\
-\frac{g^{\prime}v_{a}}{\sqrt 2}& 0 & \frac{g v_{a}}{\sqrt 2} & 0 
& 0 & 0 & -fv_{u} & 0
\end{array} \right). \nonumber \\
\label{majorana-neutralino}
\eea
An approximate expression for the tree level Majorana neutrino mass
is given by \cite{Chakraborty} 
\bea
\label{majorana-mass-tree-level}
(m_\nu)_{\rm Tree} &\simeq&-v^{2}\frac{\left[g \lambda_{T} v^{2}
(M_{2}^{D}-M_{1}^{D})\sin\beta\right]^{2}}{\left[M_{1}\alpha^{2}
+M_{2}\delta^{2}\right]},
\eea
where
\bea
\alpha&=&\frac{2 M_{1}^{D} M_{2}^{D}\gamma\tan\beta}
{g\tan\theta_{w}}
+\sqrt 2  v^{2}\lambda_{S}\tan\beta(M_{1}^{D}\sin^{2}\beta+
M_{2}^{D}\cos^{2}\beta),\nonumber \\
\delta&=&\sqrt2  M_{1}^{D}v^{2}\lambda_{T}\tan\beta,
\eea
and $\gamma$ has been defined earlier.
Note that the neutrino Yukawa coupling $f$ does not arise in this expression 
because of our choice in eq.~(\ref{MR-lambda}). Therefore, it is obvious from 
eq.~(\ref{majorana-mass-tree-level}) that in order to obtain a small tree level 
Majorana neutrino mass, we either require a small $\lambda_T$ or nearly 
degenerate Dirac gaugino masses\footnote{A detailed discussion on how to 
fit the light neutrino masses and mixing in this model can be found in 
\cite{Chakraborty}.}. In this work we are interested in the sterile 
neutrino which might play the role of keV dark matter. From the $8\times 8$ 
neutralino mass matrix, the sterile neutrino mass can be approximated as
\bea
M_N^R\simeq M_1\frac{2 f^2 \tan^2 \beta}{g^{\prime 2}}.
\label{sterile-mass}
\eea
For a wide range of parameters the active-sterile mixing can also be 
estimated as
\bea
\theta_{14}^2\simeq \frac{(m_{\nu})_{\rm Tree}}{M_N^R}.
\label{mixing}
\eea
\begin{figure}[htb]
\begin{center}
\includegraphics[height=3.5in,width=3.5in]{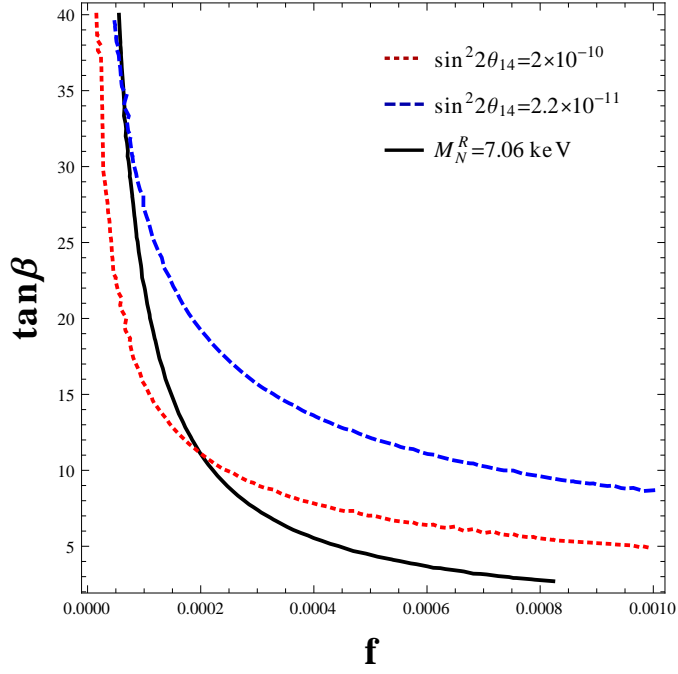} 
\caption{\label{fig:tanb_f}
The contour in the black thick line represents a sterile neutrino mass of 
7 keV. Contours in red (dotted) and blue (dashed) colours show active-sterile 
mixing $2.2\times 10^{-11}$ and $2\times 10^{-10}$ respectively.}
\end{center}
\end{figure}
In figure~(\ref{fig:tanb_f}), we show in the ($f - \tan\beta$) plane the contour 
of sterile neutrino mass fixed at 7.06 keV and also two different contours of 
$\sin^2 2\theta_{14}$, fixed at the lower and upper limit at $2.2\times 10^{-11}$ 
and $2\times 10^{-10}$ respectively. We have chosen the gravitino mass, $m_{3/2}$ 
to be 10 GeV and $M_1^D = 900 ~{\rm GeV}$, keeping a degeneracy between the Dirac 
gaugino masses, $\epsilon \equiv (M_2^D -M_1^D) = 10^{-4} ~{\rm GeV}$. We have 
also fixed $\mu_u = 750 ~{\rm GeV}$, $\lambda_S = 1.1$, $v_S = -0.1 ~{\rm GeV}$ 
and $v_{T} = 0.1 ~{\rm GeV}$. 

The sterile neutrino mass contour can be easily explained by looking at 
eq.~(\ref{sterile-mass}). Similarly from eq.~(\ref{majorana-mass-tree-level}), 
eq.~(\ref{sterile-mass}) and eq.~(\ref{mixing}), it is straightforward to show 
that $\sin^2 2\theta_{14}$ goes as $\frac{1}{1+\tan^2 \beta}$. This means that 
for smaller $\tan\beta$ one would expect larger mixing angle for fixed values 
of other parameters. This is also evident from figure~\ref{fig:tanb_f}. Furthermore, 
for larger Dirac gaugino masses, the active neutrino mass gets reduced 
(see eq.~(\ref{majorana-mass-tree-level})), which also implies a reduction 
in the active-sterile mixing. 

\begin{figure}[htb]
\begin{center}
\includegraphics[height=3.5in,width=3.5in]{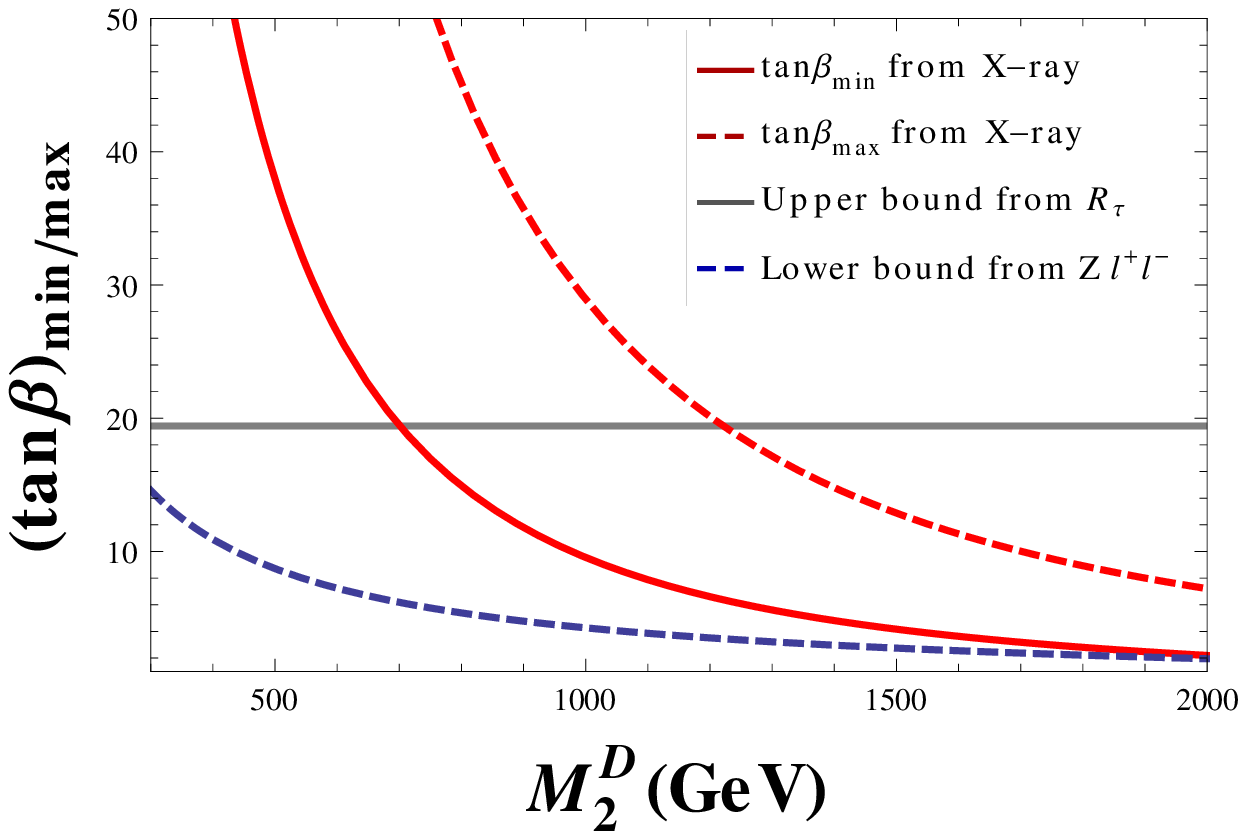} 
\caption{\label{fig:M1D_tanb}
Showing the lower and upper limits of $\tan\beta$ from X-ray analysis 
as a function of $M_2^D$ for $\mu_u$ = 700 GeV, $m_{3/2}$ = 10 GeV and 
$\epsilon = 10^{-4}$ GeV.}
\end{center}
\end{figure}
Looking at figure~\ref{fig:tanb_f}, we observe that the largest value of the 
active-sterile mixing, required to explain the observed photon line flux at 
an energy $E \approx$ 3.5 keV, corresponds to the minimum value of $\tan\beta$. 
In fact, for this particular case shown in figure~\ref{fig:tanb_f}, $(\tan\beta)_
{\rm min} \approx 11.3$. Similarly the smallest active-sterile mixing 
($\sin^2 2\theta_{14} = 2.2 \times 10^{-11}$) provides the maximum allowed 
value of $\tan\beta$, which in this case turns out to be $(\tan\beta)_{\rm max} 
\approx 33$. In order to obtain an analytical relationship between the lower 
limit of $\tan\beta$ and $M_2^D$, we can solve for $\tan\beta$ using 
eq.~(\ref{mixing}), with $\sin^2 2\theta_{14}=2\times 10^{-10}$ and 
$M_N^R = 7.06 ~{\rm keV}$. This gives rise to
\bea
(\tan^2\beta)_{\rm min} &=& \dfrac{4 v^2 \{g \lambda_T v^2 (M^D_2 - M^D_1) \}^2}
{(1.4 \times 10^{-15} ~{\rm GeV}) [M_1 {\alpha^\prime}^2 + M_2 {\delta^\prime}^2]} -1, 
\nonumber \\
\label{eq:tanb_low_high}
\eea 
where 
\bea
\alpha^\prime &\simeq& \dfrac{2 (M_2^D)^2 \mu_u}{g^\prime} + \sqrt{2} v^2 \lambda_s M^D_2, 
\nonumber \\
\delta^\prime &\simeq& \sqrt{2} M^D_2 v^2 \lambda_T. 
\eea
In a similar way an analytical expression for the upper limit of $\tan\beta$ 
can also be derived.

Figure~\ref{fig:M1D_tanb} shows the lower and upper limits of $\tan\beta$ as 
a function of $M_2^D$, for $\mu_u$ = 700 GeV, $m_{3/2}$ = 10 GeV and 
$\epsilon = 10^{-4}$ GeV. We have fixed $\lambda_S$ at the previously mentioned 
value. The horizontal grey line shows the upper limit on $\tan\beta$ arising 
from the contribution of the leptonic Yukawa coupling, $f_{\tau} \equiv 
\lambda_{133}$ to the ratio $R_\tau \equiv \Gamma(\tau \rightarrow e 
{\bar \nu}_e \nu_\tau)/\Gamma(\tau \rightarrow \mu {\bar \nu}_\mu \nu_\tau)$. The 
resulting constraint is $f_{\tau}<0.07\left(\frac{m_{{\tilde \tau}_R}}
{100~{\rm GeV}}\right)$ \cite{Kumar-1} and considering stau mass, close to 
$280 ~{\rm GeV}$, translates into an upper limit on $\tan\beta \approx 19$.
For higher stau mass this upper limit on $\tan\beta$ gets relaxed. 
The blue dashed line shows the lower bound on $\tan\beta$, as a function 
of $M_2^D$, arising from the precision measurements of the deviations in 
the couplings of the Z boson to charged leptons \cite{Kumar-1}. 

We infer from the above discussions, that in a large region of the parameter 
space, the lower limit on $\tan\beta$, satisfying the estimated mass and mixing 
of the sterile neutrino dark matter particle coming from the recent observation 
of an X-ray line signal at energy 3.5 keV is stronger than the lower limit on 
$\tan\beta$ coming from the electroweak precision measurements. On the other 
hand, the upper limit on $\tan\beta$ coming from the X-ray observations becomes 
stronger than the upper limit arising from the $\tau$ Yukawa coupling contribution 
to $R_\tau$ only for higher values of $M^D_2$ as shown in figure~\ref{fig:M1D_tanb} 
for specific choices of $\mu_u$ and $m_{3/2}$. Combining these lower and upper 
limits on $\tan\beta$ from X-ray observations and measurement of $R_\tau$, we 
can find  a range of $M^D_2$ that is allowed. For smaller values of $\mu_u$ and 
$m_{3/2}$, the upper and lower limits of $M^D_2$ shift to higher values 
(see eq.~(\ref{eq:tanb_low_high})).

We also observe from figure~\ref{fig:tanb_f} that the allowed values of $f$ is of 
the order of $10^{-4}$. Such a small value of $f$, implies negligible extra 
contribution to the tree level Higgs boson mass. Therefore, to elevate the Higgs 
boson mass to $125$ GeV, we have to rely on the loop corrections. Sizable 
radiative corrections are obtained if $\lambda_S$, $\lambda_T$ are large 
($\mathcal O(1)$) and this would imply nearly degenerate Dirac gaugino masses 
($\epsilon \sim 10^{-4}$ GeV) in order to have the active-sterile mixing 
$\sin^2 2\theta_{14} \sim 10^{-11}$ and a tree level active neutrino mass 
$\lsim$ 0.05 eV.
The other case, which can relax this strong degeneracy between Dirac gaugino 
masses, corresponds to the case of small $\lambda_S$, $\lambda_T\sim\mathcal 
O(10^{-4})$, which implies multi-TeV stop to fit the Higgs boson mass. Therefore, 
this model provides a very interesting possibility where we can connect the 
Higgs sector with the neutrino sector (both active as well as sterile neutrino).  

\section{Right handed neutrino as a keV warm dark matter}
To accommodate sterile neutrino as a warm dark matter candidate, it 
is very important to make sure that the active sterile mixing is
very small~\cite{Abazajian-1,Shaposhnikov,Shaposhnikov-1,Biermann,Dahle,
Roy} and within the valid range of different X-ray experiments. 
A rough bound on the active-sterile mixing can be parametrised as
\cite{Boyarsky-1}
\bea
\theta_{14}^{2}\leq 1.8\times 10^{-5}\Big(\frac{1 \rm{keV}}{M_N^R}\Big)^5.
\eea
Along with the strict bound coming from different X-ray experiments, 
the keV sterile neutrino must produce the correct relic density $\Omega_N 
h^2\sim 0.1$, in order to identify itself with the warm dark matter. 
An approximate formula for the relic density of sterile neutrinos, produced 
in the early universe with negligible lepton asymmetry via non-resonant 
oscillations with active neutrinos, known as the Dodelson-Widrow (DW) mechanism 
\cite{Dodelson} can be written as \cite{Fuller}
\bea
\Omega_N h^2\sim 0.3 \Big(\frac{\sin^2 2\theta_{14}}{10^{-10}}\Big)
\Big(\frac{M_N^R}{100 ~\rm{keV}}\Big)^2, 
\eea
where $\Omega_N$ is the ratio of the sterile neutrino density to the 
critical density of the Universe and $h=0.673$. 

Different experimental observations have also put lower limits on the mass 
of the keV warm dark matter. A very robust bound for fermionic dark matter 
particles comes from Pauli exclusion principle. By claiming the maximal 
(Fermi) velocity of the degenerate fermionic gas in the dwarf spheroidal 
galaxies is less compared to the escape velocity, translates into a lower 
bound on the sterile neutrino dark matter mass, i.e $M_N^R>0.41$ keV 
\cite{Tremaine-Gunn,Boyarsky-mass-bound}. Model dependent bounds on the 
mass of the warm dark matter are much more stringent and obtained from 
analysing Lyman-$\alpha$ experiment \cite{Seljak, Lesgourgues}. 

\begin{figure}[htb]
\begin{center}
\includegraphics[height=3.5in,width=3.5in]{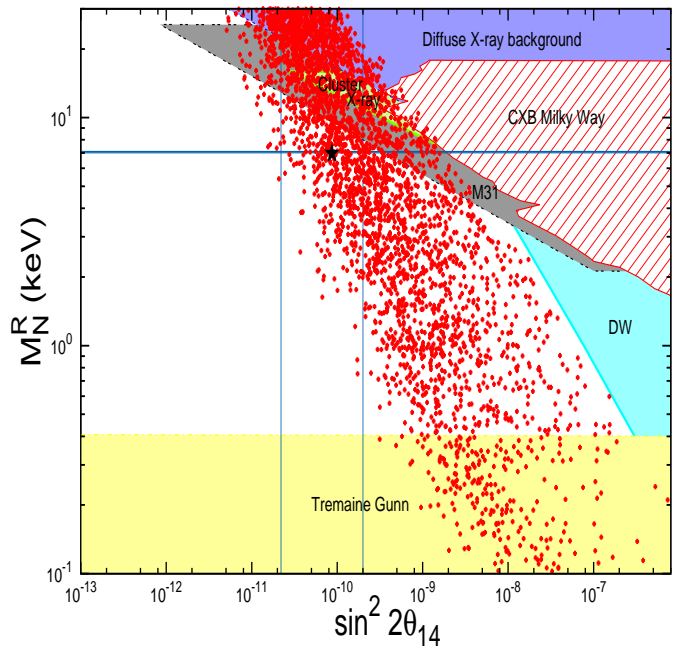} 
\caption{\label{fig:light_stop}
The red (grey) points in the mass-mixing plane are obtained by scanning the parameter 
space as mentioned in the text. The yellow (light) region is ruled out from the 
Tremaine Gunn bound \cite{Tremaine-Gunn,Boyarsky-mass-bound}. Cosmic X-ray 
background (CXB) rules out the region in red stripes \cite{X-ray-bound}. 
Constraints from M31, observed by Chandra rules out the region in grey 
\cite{Yuksel-Watson}. The blue region is ruled out from the diffuse X-ray background 
observations \cite{X-ray-bound}. XMM-Newton observations from Coma and Virgo 
clusters rule out the region in green. The light blue line represents the 100 
\% relic density of the sterile neutrino dark matter, produced via DW mechanism. 
The light blue region above this line leads to over abundance of the sterile 
neutrino warm dark matter. Finally, the black star represents the central value of 
the mass and active-sterile mixing, from the 3.5 keV X-ray line observation.} 
\end{center}
\end{figure}

In figure~\ref{fig:light_stop} we present a scatter plot by scanning the parameter
space of our model and also show the compatibility of those points with the
current experimental findings. The red circles are the points obtained by varying 
the parameters as $500~\rm{GeV}<M_1^D<1.2~\rm{TeV}$, $10^{-5}<f<10^{-3}$, 
$2.7<\tan\beta<17$, $400~\rm{GeV}<m_{\tilde t_1,\tilde t_2}<1.2 ~{\rm TeV}$, keeping 
$\epsilon \equiv (M_2^D-M_1^D)\sim 10^{-4}~\rm{GeV}$. $\mu_u$ and $\lambda_S$ are 
fixed at 750 GeV and 1.1 respectively ($\lambda_T = \lambda_S \tan\theta_W \sim 0.6$). 
All these points respect a Higgs boson mass in between 124.4 GeV and 126.2 GeV 
avoiding any tachyonic scalar states.

Similar plot can also be generated where $\lambda_T\sim 10^{-5}$. Therefore, 
to fit the Higgs boson mass in that case, one requires $m_{\tilde t} > 5~\rm{TeV}$. 
However, the degeneracy between $M_1^D$ and $M_2^D$ is somewhat lifted where 
$\epsilon \gsim 1~\rm{GeV}$. 

The horizontal yellow band in figure\ref{fig:light_stop} is ruled out by the Tremaine 
Gunn bound, which implies $M_N^R<0.4~\rm{keV}$ \cite{Tremaine-Gunn,
Boyarsky-mass-bound}. The blue region is excluded by taking into consideration 
the diffuse X-ray background \cite{X-ray-bound}. Cluster X-ray bound rules out 
a region in the mass-mixing plane by taking into consideration XMM-Newton 
observations from the Coma and Virgo clusters \cite{Cluster-x-ray}. Constraints 
from the cosmic X-ray background (CXB) rules out the region in red stripes 
\cite{X-ray-bound}. Chandra observation of M31 \cite{Yuksel-Watson} rules 
out the region in grey. The light blue line corresponds to the correct relic density
provided by the sterile neutrino warm dark matter via DW mechanism. The light
blue region above this line marked as DW is ruled out because of the over abundance of 
sterile neutrino dark matter. The horizontal and vertical lines show the region 
in the mass and mixing plane consistent with the observed  3.5 keV X-ray line with more 
than 3$\sigma$ significance. The black star corresponds to the best fit point. 
It is clearly evident from this figure that such 
a small mixing is completely in conflict with the DW production mechanism of 
sterile neutrinos. However, resonant production of sterile neutrinos in the 
presence of a lepton asymmetry in primordial plasma can be very important and 
produce correct relic abundance of the keV sterile neutrinos \cite{Fuller,Shi}. 
Recent studies have shown that a cosmological lepton asymmetry $L \sim 
\mathcal O(10^{-3})$ is capable of producing correct relic density of 0.119 
\cite{Abazajian}. It was shown in \cite{Volkas, Shi-1, Foot-Volkas, Foot, Kishimoto,
Dolgov,Dolgov-1,Maalampi} 
that active-sterile neutrino oscillations can themselves create a cosmological 
lepton number of this magnitude, assuming that the number of sterile neutrinos 
is negligible to start with. Such a possibility can be easily conceived in 
our model to generate a large lepton asymmetry.  

Let us note in passing that sterile neutrino production in non-standard 
cosmology with low reheating temperature ($\sim$ a few MeV) has also been 
discussed in the literature \cite{Pascoli,Yaguna,Osoba}. If the universe has 
undergone inflation and was never reheated to a temperature above a few MeV 
then the relic abundance of the sterile neutrinos can be written as
\bea
\Omega_N h^2 = 10^{-7} d_{\alpha}\Big(\frac{\sin^2 2\theta_{14}}{10^{-10}}\Big)
\Big(\frac{M_N^R}{10 ~\rm{keV}}\Big)\Big(\frac{T_R}{5 ~{\rm MeV}}\Big)^3,
\eea 
where $d_{\alpha}=1.13$, assuming that the sterile neutrino couples only with 
$\nu_e$ as in our case. It is obvious from the above expression that for allowed 
values of $\sin^2 2\theta_{14}$ and $M_N^R$ (from the recent X-ray observation) 
this production mechanism will give rise to severe under abundance of sterile 
neutrinos.

In our model sterile neutrinos can also be produced non-thermally via the decay 
of heavier scalar particles. However, a quantitative estimate of the relic 
density requires a thorough investigation and we postpone the discussion of 
this  method of production for a future work \cite{Chakraborty-1}.
\begin{figure}[htb]
\begin{center}
\includegraphics[height=3.5in,width=3.5in]{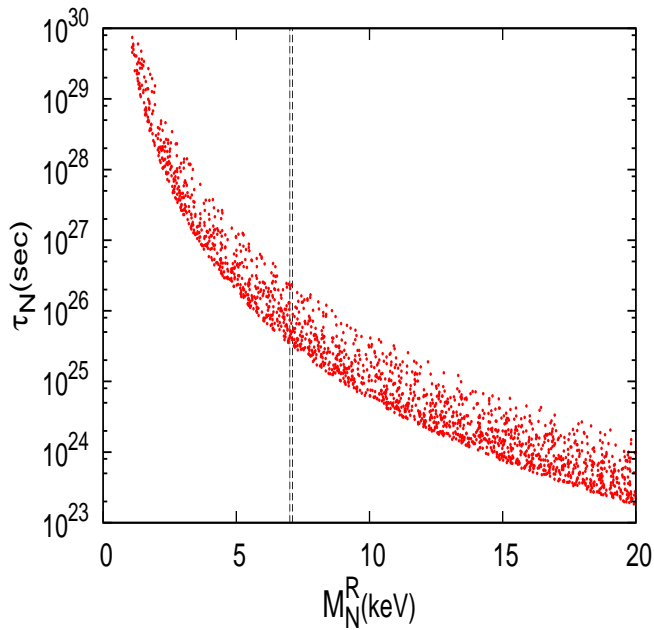}
\caption{\label{fig:sterile_lifetime}
The plot shows the sterile neutrino lifetime as a function of 
the sterile neutrino mass. The black vertical line represents
the 7 keV mass of the sterile neutrino. The red points represent the 
total life time of the sterile neutrino.} 
\end{center}
\end{figure}

\subsection{Sterile neutrino decay}
The most dominant decay mode of the sterile neutrino is $N\rightarrow 3\nu$. 
The corresponding decay rate for this process is given by \cite{Abazajian-1}
\bea
\Gamma_{3\nu}=8.7\times 10^{-31} {~\rm{sec}}^{-1}
\Big(\frac{\sin^2 2\theta_{14}}{10^{-10}}\Big)
\Big(\frac{M_N^R}{1 ~\rm{keV}}\Big)^5.
\eea
The principal radiative decay mode of the sterile neutrino which 
is of concern here is $N\rightarrow \nu\gamma$ and the decay width
is 
\bea
\Gamma_{\nu\gamma} = 1.38 \times 10^{-32} ~{\rm{sec}}^{-1} 
\Big(\frac{\sin^2 2\theta_{14}}{10^{-10}}\Big)
\Big(\frac{M_N^R}{1 ~\rm{keV}}\Big)^5.
\label{sterile_decay}
\eea
This decay produces a monochromatic photon line at $E_{\gamma}=
\frac{M_N^R}{2}$. 
From figure~(\ref{fig:sterile_lifetime}) we can see that the lifetime of 
the sterile neutrino is much larger than the age of the universe. 
\section{Gravitino cosmology}
As mentioned earlier, the gravitino mass is the order parameter of R-breaking. 
If the mass is around a few GeV, it can be a candidate for cold dark matter 
\cite{Covi}. In our scenario, the gravitino is an unstable particle and decays 
to an active/sterile neutrino and a monochromatic photon. The tree level decay mode 
into an active neutrino final state $\tilde G\rightarrow \gamma\nu_e$ 
is suppressed by the very small mixing $U_{\tilde b\nu_e}$ ($\sim 10^{-7}$) 
between the bino and active neutrino $\nu_e$ \cite{Takayama}. Interestingly, in 
our model the most dominant decay mode of gravitino is into a photon and a sterile 
neutrino ($\tilde G\rightarrow N\gamma$) and the decay width is given as
\bea
\Gamma_{\tilde G\rightarrow N\gamma}\sim \frac{|U_{\tilde b N}|^{2}m_{3/2}^{3}}
{32\pi M_{P}^{2}},
\eea
where $U_{\tilde b N}$ is the bino sterile neutrino mixing angle. Because of the
presence of the term $M_R {\hat N}^c {\hat S}$ in the superpotential and the 
bino Dirac mass term in the Lagrangian, the tree level bino sterile neutrino 
mixing is not strongly suppressed ($\sim 10^{-2}$).  

For the sake of completeness, let us mention that at the one loop level the
decay $\tilde G\rightarrow \gamma\nu_e$ occurs \cite{Masiero,Lola,Lola-1,
Ghosh-Zhang} via trilinear R-parity violating coupling $\lambda_{133}^{\prime}$ 
which we have identified with the bottom Yukawa coupling. We have checked that 
this process is also suppressed compared to the tree level decay $\tilde 
G\rightarrow N\gamma$. The one-loop contribution to the decay $\tilde G\rightarrow 
N\gamma$ is negligible because of small active-sterile mixing.


Taking into account the most dominant decay mode of the gravitino in the sterile
neutrino plus photon final state, for a 10 GeV gravitino mass, the lifetime is close 
to $10^{15}$ sec. Therefore, to satisfy the experimental constraints coming from the 
diffuse photon background, one has to consider a scenario where the gravitino density 
is very much diluted. In order to provide a quantitative analysis we note that 
for a gravitino of mass 10 GeV the limit on the diffuse photon flux is around 
$6.89\times 10^{-7}\rm{GeV}\rm {cm^{-2}}\rm{sec^{-1}}$ \cite{Yuksel}. This can 
be translated into a bound on the gravitino relic density and we find
\bea
\Omega_{3/2}h^2< 4.34\times 10^{-13} \left(\frac{10^{-2}}{U_{\tilde bN}}\right)^2,
\eea
for a 10 GeV gravitino. Note that this bound depends strongly on the mass of
the gravitino and will get relaxed for a smaller gravitino mass.
To satisfy such a strong bound on the gravitino relic density,
one must account for a very low reheating temperature. If the reheating
temperature is above the SUSY scale, the gravitino relic density would be
too large \cite{Buchmuller}. Therefore, the reheating temperature must lie 
much below the SUSY threshold.

Following \cite{Gregoire}, we see that if the reheating temperature is below 
the SUSY threshold, the gravitinos are produced by thermal scattering with 
neutrinos and bottom quarks. Using the results of \cite{Gregoire} for 
production of gravitinos, we obtain an upper bound on the reheating temperature 
for a 10 GeV gravitino as
\bea
T_{R}&<&127 \left(\frac{v_a}{30\rm{GeV}}\right)^{2/7}
\left(\frac{m_{\tilde b}}{500\rm{GeV}}\right)^{4/7}
\left(\frac{10^{-2}}{U_{\tilde b N}}\right)^{2/7}\rm{GeV}. \nonumber \\
\eea
Such a low reheating temperature might have important implications
in the context of different baryogenesis and leptogenesis scenarios.
\section{Conclusion}
Recent observation of a weak X-ray line around $E_{\gamma}=3.5~\rm{keV}$
by XMM-Newton telescope coming from Andromeda galaxy and various 
galaxy clusters have been studied in the light of a $U(1)_{R^-}$lepton
number model, with a single right handed neutrino. We have shown explicitly
that a sterile neutrino of mass about 7 keV and with appropriate
active-sterile mixing can easily be obtained in our model. We briefly 
mention different production mechanisms of the sterile neutrino.

Allowed ranges of the mass and mixing helped us to put bounds on 
$\tan\beta$ as a function of the Dirac wino mass $M_2^D$. Combining these
bounds with the limits coming from the measurements of the $\tau$ Yukawa 
coupling contribution to the ratio $R_\tau \equiv \Gamma(\tau \rightarrow 
e {\bar \nu}_e \nu_\tau)/\Gamma(\tau \rightarrow \mu {\bar \nu}_\mu \nu_\tau)$, 
one obtains strong upper and lower bounds on $M^D_2$. 

In addition, we have also discussed the Higgs sector briefly and pointed out 
different possibilities to have a Higgs boson mass around 125 GeV. Finally, 
gravitino is the LSP in our model with a mass about a few GeV and gravitino 
mass is the order parameter of R-symmetry breaking. The gravitino can decay 
into a photon plus active or sterile neutrino. Therefore, we have also 
presented a short discussion on the cosmological implications of the gravitino. 
We have taken into account the most robust constraint coming from the diffuse 
photon background, which readily puts a very stringent bound on the gravitino 
relic density. This eventually imposes an upper limit ($\lsim$ 130 GeV) on the 
reheating temperature of the universe.

\begin{acknowledgments}
We thank Kevork N. Abazajian and George G Raffelt for helpful discussions. 
S.C would also like to thank the Council of Scientific and Industrial Research 
(CSIR), Government of India for financial support obtained as a Senior Research 
Fellow.
\end{acknowledgments}

\end{document}